\documentclass[pre,
   superscriptaddress,
twocolumn,
   eqsecnum,
   showpacs,
   preprintnumbers,
   byrevtex]{revtex4}

\newcommand{\bq}{\begin{equation}}
\newcommand{\ba}{\begin{eqnarray}}
\newcommand{\eq}{\end{equation}}
\newcommand{\ea}{\end{eqnarray}}
\newcommand {\bPsi}{{\bar \Psi}}
\newcommand {\bpsi}{{\bar \psi}}
\newcommand{\sech} {{\rm sech}}

\newcommand{\pstar}{\mbox{$\psi^{\ast}$}}

\usepackage{amsfonts}
\usepackage{amssymb}
\usepackage{amsmath}
\usepackage{color}
\usepackage{graphicx}
\usepackage[pdftex,colorlinks=true]{hyperref}
\begin{document}

\preprint{LA-UR 10-04795}

\title{Solitary waves  in the Nonlinear Dirac Equation with arbitrary nonlinearity}

\author{Fred Cooper} 
\email{fcooper@lanl.gov}
\affiliation{Santa Fe Institute, Santa Fe, NM 87501, USA}
\affiliation{Theoretical Division and Center for Nonlinear Studies, 
Los Alamos National Laboratory, Los Alamos, New Mexico 87545, USA}

\author{Avinash Khare} \email{khare@iopb.res.in}
\affiliation{ Institute of Physics, Bhubaneswar 751005, India}

\author{Bogdan Mihaila}
\email{bmihaila@lanl.gov}
\affiliation{Materials Science and Technology Division, 
Los Alamos National Laboratory, Los Alamos, New Mexico 87545, USA} 

\author{Avadh Saxena}
 \email{avadh@lanl.gov}
\affiliation{Theoretical Division and Center for Nonlinear Studies, 
Los Alamos National Laboratory, Los Alamos, New Mexico 87545, USA}

\begin{abstract}
We consider the nonlinear Dirac equations (NLDE's)  in 1+1 dimension with 
scalar-scalar  self interaction 
$\frac{ g^2}{k+1} ( {\bar \Psi} \Psi)^{k+1}$, as well as a vector-vector self interaction 
$ \frac{ g^2}{k+1} ( {\bar \Psi} \gamma_\mu \Psi \bPsi \gamma^\mu \Psi)^{\frac{1}{2}(k+1)}$.
We find the exact analytic form for  solitary waves for arbitrary~$k$ and find that they are a generalization of the exact solutions
for the nonlinear Schr\"odinger equation (NLSE) and reduce to these solutions in a well defined nonrelativistic limit.  We perform  the  nonrelativistic reduction and find the $1/2m$ correction to the NLSE, valid when $|\omega-m |\ll  2m$, where $\omega$ is the frequency  of the solitary wave in the rest frame.   We discuss the  stability and blowup of solitary waves assuming the modified NLSE is valid and find that they should be stable for $k < 2$. 
\end{abstract}

\pacs{PACS: 11.15.Kc, 03.70.+k, 0570.Ln.,11.10.-s}

\maketitle
\section{Introduction}

Beyond the usual applications in field theory, the nonlinear Dirac equation (NLDE) 
also emerges in various condensed matter applications.  An important example 
being the Bose-Einstein condensate (BEC) in a honeycomb optical lattice in the long 
wavelength, mean field limit \cite{haddad}.  The multi-component BEC order 
parameter has an exact spinor structure and serves as  the bosonic analog to the 
relativistic electrons in graphene. 

Classical solutions of nonlinear field equations have a long history as a 
model of extended particles~\cite{ref:extended,soler1}. The stability of such 
solutions in 3+1 dimensions was studied in detail by 
Derrick \cite{ref:derrick}.  He showed that the classical solutions of the
self-interacting scalar theories (with both polynomial and non-polynomial 
interactions)  were unstable to scale transformations. However he was not 
able to make any conclusive statements about the spinor theories.  In 1970, 
Soler \cite{soler1} proposed that the self-interacting 4-Fermi theory was an 
interesting model for extended fermions.  Later, Strauss and Vasquez 
\cite{ref:strauss}  were able to study the stability of this model under 
dilatation and found the domain of stability for the Soler solutions. 
Solitary waves in the 1+1 dimensional nonlinear Dirac equation have 
been studied \cite{ref:Lee,ref:Nogami} in the past in case 
the nonlinearity parameter $k=1$,  i.e. massive Gross-Neveu~\cite{ref:GN} (with
$N=1$, i.e. just one localized fermion) and massive Thirring~\cite{ref:TM} 
models). In those studies it was found that these equations have solitary wave 
solutions for both scalar-scalar (S-S) and vector-vector (V-V) interactions. 
The interaction between solitary waves of different initial charge was 
studied in detail  for the S-S case when $k=1$ 
in the work of Alvarez and Carreras \cite{ref:numerical} 
by Lorentz boosting the static solutions and allowing them to scatter. Stability
of the $k=1$ problem was also studied by Bogolubsky  \cite{ref:bogol}, who 
found using a variational method that preserved charge, that the frequencies  
$\omega < 1/\sqrt{2} $ should be unstable. However, subsequent numerical work 
by Alvarez and Soler \cite{ref:soler} showed that this result was incorrect 
(i.e. the solitary waves were numerically stable). Further analytic work on 
stability for the S-S model using the Shatah-Strauss formalism  
\cite{Shatah} by Blanchard \emph{et al.} \cite{Blanchard}  turned out to give 
inconclusive results in that they could not prove that the solutions to the 
Dirac equation were minima of the variational energy functional.  Thus the 
domain of stability of solutions to self interacting 4-Fermi theories is still 
an open question. 
 
In this paper we generalize the work of Lee, Kuo, and Gavrielides 
\cite{ref:Lee} to arbitrary~$k$ and find exact solutions for all $k$. The 
paper is organized as follows: In Sec. II we  find rest-frame solitary wave 
solutions of the form $\Psi(x,t) = e^{-i\omega t} \psi(x)$, for both the case 
of the S-S and V-V interactions. We calculate the rest frame  frequency, 
$\omega$, and the energy, $H$, of a solitary wave  of charge $Q$, as a function
 of the parameters $k$ and~$g$.  We find the range of $k$ and~$g$ values 
for which $\omega$ and~$H$ are in the range 
$ 0 < \left(\begin{array}{c} \omega \\  H  \end{array}\right) < m$.  
In Sec. III we derive the nonrelativistic limit of the NLDE and find the 
leading term which is the nonlinear Schr\"odinger equation with corrections of 
the order of $1/2m$. Our derivation agrees with the heuristic result for $k=1$ 
for modification of the NLSE found earlier by  \cite{ref:Toyama}.  We find 
that the correction term has the same magnitude but opposite sign for the V-V 
as compared to the S-S case and find that the expansion is always valid 
whenever $|\omega-m| \ll 2m$. In the V-V case, the NLDE solutions are 
numerically quite close to those of the NLSE for all values of $\omega$. 
However for the S-S case, when we depart from the domain of validity of the 
non-relativistic reduction,  the solitary wave solutions depart dramatically 
from the NLSE limit and become {\it double humped}.  We plot the crossover 
to this regime as a function of the nonlinearity parameter $k$.  In section IV we 
first discuss stability of solitary waves in the NLSE  using an auxiliary 
Lagrangian for the static solutions.  We find that the criteria for stability 
is $0<k<2$ and that identical results are obtained using stability against 
scale transformations (Derrick's theorem \cite{ref:derrick}). However the 
scale transformation argument leads to the conclusion that there should be 
unstable solitary waves in the NLDE for $k>1$ which violates continuity 
argument to the nonrelativistic regime. It also led to contradictions with 
numerical experiments at $k=1$.  We then discuss the stability question in 
the modified NLSE (mNLSE) and show that it is essentially the same as for the 
NLSE.  In section V we discuss how to obtain information about self-focusing in  
case $k=2$ and~$k>2$ for both the NLSE and mNLSE assuming that the time 
dependent solitons are self-similar generalizations of the exact solution of 
the NLSE. We find that the correction terms in the mNLSE eventually 
dominate at late times during self-focusing and so the approximation breaks 
down during the late stages of self-focusing.  

 We conclude with a summary of our main findings as well as a discussion about
the possible future directions for settling issues of stability using various
approaches including numerical methods.  


\section{Solitary wave solutions}

We are interested in solitary wave solution of the NLDE given by 
\bq
(i \gamma^{\mu} \partial_{\mu} - m) \Psi +g^2 (\bPsi  \Psi)^{k} \Psi = 0 \>,
\eq
for the scalar-scalar  interaction
and 
\bq
(i \gamma^{\mu} \partial_{\mu} - m) \Psi +g^2  \gamma^\mu \Psi (\bar \Psi \gamma_\mu \Psi) ( \bar \Psi \gamma_\mu \Psi 
\bPsi \gamma^\mu \Psi )^{\frac{1}{2}(k-1)} = 0 \>,
\eq
for the vector-vector interaction.
These equations  can be derived in a standard fashion from the Lagrangian
\bq
L =  \overline{\Psi} \left(i \gamma^{\mu} \partial_{\mu} - m \right) \Psi 
+  L_I  \>.
\eq
For scalar-scalar interactions, we have 
\bq
L_I  =   \frac{g^2}{k+1} (\overline{\Psi} \Psi)^{k+1} \>, 
\label{eq:t112}
\eq
whereas for vector-vector interactions  we have instead 
\bq
L_I  =   \frac{ g^2}{k+1} ( {\bar \Psi} \gamma_\mu \Psi 
\bPsi \gamma^\mu \Psi)^{\frac{1}{2}(k+1)} \>. \label{eq:vector}
\eq 
Note that in the above equations, $g^2$ is the dimensional coupling constant, i.e.  $g^2=G^2 m^{1-k}$, where $G$ is dimensionless. 
The $\gamma$ matrices in 2 dimensions in our convention satisfy
\bq
\{\gamma_\mu,\gamma_\nu \}_+ = 2 g_{\mu \nu} ; ~~ g_{\mu \nu} =\left(\begin{array}{cc} 1 & 0 \\0 & -1 \\ \end{array}\right).
\eq
We are looking for  solitary wave solutions where the  field $\Psi$ goes to zero at infinity.    It is sufficient to go into 
the rest frame, since the theory is Lorentz invariant and the moving solution can be obtained by a Lorentz boost.
In the rest frame we have that 
\bq
\Psi(x,t) = e^{-i\omega t} \psi(x) \>.
\eq
We are interested in bound state solutions that correspond to positive frequency in the rest frame less than the mass parameter $m$, i.e. $0 \leq \omega < m$.  For these bound state solutions one requires that the energy of the solitary wave $H$ obeys
$0 \leq  H < m$. 
Choosing the representation  $\gamma_0 = \sigma_3$, $i \gamma_1= \sigma_1$, where the $\sigma_i$ are the standard Pauli spin matrices, 
we obtain 
\bq
i  \sigma_3  \partial_t \Psi + \sigma_x \partial_x \Psi - m \Psi - V_I \Psi =0,  \label{NLDEa}
\eq
where $ V_I = - \frac{\partial L_I}{\partial {\bar \Psi} }$. 
Defining the matrix, 
\ba
\psi(x) &&=    \left(\begin{array}{c}u \\v\end{array}\right)  = R(x) \left(\begin{array}{c}\cos \theta \\ \sin \theta \end{array}\right) \>,
\ea
we obtain the following equations for $u$ and~$v$.
For scalar-scalar interactions, we find: 
\ba
&& \frac{du}{dx} + (m+\omega ) v - g^2(u^2-v^2)^k v=0 \>, \nonumber \\
&&\frac{dv}{dx} + (m-\omega ) u - g^2(u^2-v^2)^k u=0 \>. \nonumber \\
\ea
For the vector-vector case one has instead: 
\ba
&& \frac{du}{dx} + (m+\omega ) v +g^2(u^2+v^2)^k  v=0 \>, \nonumber \\
&&\frac{dv}{dx} + (m-\omega ) u- g^2(u^2+v^2)^k  u=0 \>. \nonumber \\
\ea
A first integral of these equations can be obtained using conservation of the energy-momentum tensor, 
\ba
T_{\mu \nu} = i  \bPsi \gamma_\mu \partial_\nu \Psi - g_{\mu \nu} L \>,
\nonumber \\
\partial^\mu  T_{\mu \nu} =0 \>,
\ea
which yields  for stationary solutions
\bq
T_{10} = \mathrm{constant}\>, \quad T_{11} = \mathrm{constant} \>.
\eq
For all the cases we want to study we can write 
\bq
T_{11} = \omega  \psi^\dag \psi - m  \bpsi \psi + L_I \>.
\eq
For solitary wave solutions vanishing at infinity the constant is zero and we get the useful first integral:
\bq
T_{11}= \omega  \psi^\dag \psi - m  \bpsi \psi + L_I = 0 \>. 
\label{eq:t11}
\eq
Multiplying the  equation of motion for either the scalar-scalar or vector-vector interaction  on the left by $\bpsi$  we have that:
\bq
(k+1)  L_I = - \omega  \psi^\dag \psi + m  \bpsi \psi  + \bpsi i \gamma_1 \partial_1 \psi  \>. 
\label{eq:motion}
\eq
We find from Eqs. \eqref{eq:t11} and \eqref{eq:motion} that
\bq
\omega  k  \psi^\dag \psi - m k    \bpsi \psi  + \bpsi i \gamma_1 \partial_1 \psi = 0 \>.
\eq
For  the Hamiltonian density we have
\ba
 {\cal H} &=&T_{00} =   \bpsi i \gamma_1 \partial_1 \psi+ m \bpsi \psi - L_I  
 \nonumber \\
 &\equiv&  h_1+ h_2- h_3 \>. \label{eq:hdensity}
\ea
Each of ${h_i}$ are positive definite.
From Eq. \eqref{eq:t11} and \eqref{eq:motion} one derives that
\bq
 k  L_I =   \bpsi i \gamma_1 \partial_1 \psi \>,   \label{eq:rel}
\eq
which further implies that
\bq
h_3 = \frac{1} {k}  h_1 \>.  \label{eq:relation}
\eq
In particular, for $k=1$, we obtain ${\cal H} = m \bpsi \psi$. In terms of 
$(R, \theta)$ one has 
\bq
 \bpsi i \gamma_1 \partial_1 \psi = \psi^\dag \psi \frac{d \theta} {dx} \>.
 \eq
This leads to the simple differential equation for $\theta$  for solitary waves 
 \bq
 \frac{d \theta} {dx} = -  \omega _k+ m_k \cos 2 \theta \>,
 \eq 
where $\omega _k  \equiv k \omega$ and~$m_k = k m$.
The solution is
\bq
\theta(x) =  \tan^{-1} ( \alpha \tanh \beta_{k} x) \>, \label{eq:theta}
\eq
where 
\bq
\alpha = \sqrt{ \frac{m_k - \omega _k}{m_k + \omega _k} }
=  \sqrt{ \frac{m - \omega }{m + \omega } }, \quad \beta_{k} =  \sqrt{ m_k^2 - \omega _k^2 } \>.
\eq
In what follows it is often useful to rewrite everything in terms of $\alpha$ and~$\beta$.  We have the relations:
\bq
m+\omega = \frac{\beta}{\alpha},~~ m- \omega = \alpha \beta\,,~~\beta=
\sqrt{m^2-\omega^2}\,.
\eq

\subsection{Scalar-Scalar interaction}

First let us look at the S-S interaction. 
Using Eqs.~\eqref{eq:t112}  and \eqref{eq:t11} we obtain 
\bq
\omega R^2 - mR^2 \cos 2 \theta 
+  \frac{g^2}{k+1} \left( R^2 \cos 2 \theta \right) ^{k+1}=0 \>.
\eq
Thus 
\bq
R^2 =\left[  \frac{(k+1) (m \cos 2 \theta -\omega ) }{g^2 (\cos 2 \theta)^{k+1} }\right] ^{\frac{1}{k}} \>.
\eq
We have
\bq
\frac{d \theta}{dx} = \frac{\beta_{k}^2}{\omega _k+m_k \cosh 2\beta_{k} x} 
                                = -\omega _k + m_k \cos 2 \theta \>,
\eq
so that 
\bq  \cos 2 \theta =  \frac{m_k+\omega _k \cosh 2 \beta_{k} x}{\omega _k+m_k \cosh 2 \beta_{k} x }=  \frac{m+\omega  \cosh 2 \beta_{k} x}{\omega +m \cosh 2 \beta_{k} x } \>.
\eq
One important expression is 
\bq
m \cos 2 \theta - \omega  = \frac {\beta_{k}^2} {k^2(\omega +m \cosh 2 \beta_{k} x)} \>.
\eq
Using this we get
\bq
R^2 = \frac {\omega +m \cosh 2 \beta_{k} x}{ m+\omega  \cosh 2 \beta_{k} x} 
\left[ \frac {(k+1) \beta_{k}^2} {g^2 k^2 (m+\omega  \cosh 2 \beta_{k} x)} \right]^{\frac{1}{k}} \>, 
\label{eq:RsqSS}
\eq
Using the identities:
\ba
1+ \alpha^2 \tanh^2 \beta_{k} x & = & \left(\frac {m \cosh 2 \beta_{k} x + \omega }{m+\omega } \right) \sech^2\beta_{k} x  \>,
\nonumber \\
1- \alpha^2 \tanh^2 \beta_{k} x &  = & \left(\frac {\omega \cosh 2 \beta_{k} x + m}{m+\omega } \right)  \sech^2\beta_{k} x \>,
\nonumber \\
\ea
we obtain the alternative expression
\bq
R^2 = \frac  {1+\alpha^2 \tanh^2\beta_{k} x }  {1-\alpha^2 \tanh^2\beta_{k} x } 
  \left[ \frac{ (k+1) \beta_{k}^2 \sech^2 \beta_{k} x  }{g^2 k^2 (m+\omega) ( 1-\alpha^2 \tanh^2\beta_{k} x )} \right ]^{\frac{1}{k}}  \>.
  \label{eq:Rsq}
\eq
The equation for $\omega$ in terms of $g^2$ is determined from the fact that 
the single solitary wave has charge Q,
\bq
 Q=  \int _{-\infty}^{\infty} dx \ \psi^\dag \psi = \int _{-\infty}^{\infty} dx \ R^2(x) \>. \label{eq:Q}
\eq
Thus the equation we need to solve for $\omega$ is 
\bq
Q=  \frac{1}{ \beta_{k} } \left [ \frac{(k+1) \beta_{k}^2}{g^2 k^2 (m+\omega )} \right ]^{1/(k)} I_k[\alpha^2]  \>, \label{eq:Qa}
\eq
where
\bq
I_k[\alpha^2] 
=\int^{1}_{-1} dy \, \frac{1+\alpha^2 y^2}{(1-y^2)^{\frac{1}{k}(k-1)}(1-\alpha^2 y^2)^{\frac{1}{k}(k+1)}} \>.
\eq
For $k=1$,  one obtains
\ba
I_{1}[\alpha^2]&=& \frac{2}{1-\alpha^2} \>, 
\\ \nonumber
Q  &=& \int_{-\infty}^{\infty} dx \ R^2 = \frac{4\alpha}{(1-\alpha^2)g^2}= \frac {2 \beta}{g^2 \omega} \>, \label{eq:Q1} 
\ea
with the solution
\bq
\omega = \frac{m}{\sqrt{1+Q^2 g^4/4}},
\eq
in agreement with earlier results of \cite{ref:Lee}. 
For $k=\frac{1}{2}$, we obtain 
\ba
I_{\frac{1}{2}}[\alpha^2] &=& \int_{-\infty}^{\infty} dy \  \frac{(1-y^2)(1+\alpha^2 y^2)}{(1-\alpha^2 y^2)^3} 
\\ \nonumber
&=&\frac{4}{\alpha^3} \left( \tanh^{-1}\alpha - \alpha \right) 
\ea
and  
\bq
Q = \frac{(k+1)^2 \beta_k \alpha^2 }{k^2 g^4}  ~~I_{\frac{1}{2}}.
\eq
For $k=\frac{3}{2}$, we obtain 
\bq
I_{\frac{3}{2}}[\alpha^2] =-2 K\left(\alpha^2\right)+\frac{4 \omega \left(\alpha^2\right)}{1-\alpha^2} \>,
\eq
where $K(k)$ is the complete elliptic integral of the first kind.  In general we can cast $I_k[\alpha^2] $ into the  sum of two hypergeometric functions $_2F_1$.  Letting $y=x^{\frac{1}{2}}$ we have that 
\bq
I_k[\alpha^2] = \int_0^1  dx \  \frac{x^{-\frac{1}{2}} (1+ \alpha^2 x) (1-x) ^{-\frac{1}{k}(k-1)}}{(1-\alpha^2 x)^{\frac{1}{k}(k+1)}} \>.
\eq
From the definition
\bq
 \int_0^1  dt \ \frac{t^{b-1} (1-t) ^{c-b-1}}{(1-tz)^{a}}= \frac{\Gamma(b) \Gamma(c-b)}{\Gamma(c)} \ _2F_1(a,b,c;z) \>,
 \eq
we find that
\ba
&&
I_k[\alpha^2, k] 
\\ \nonumber 
&& = B \Bigl (\frac{1}{2},\frac{1}{k}\Bigr ) \phantom{a}_2F_1\Bigl (1+\frac{1}{k},\frac{1}{2},\frac{1}{2} +\frac{1}{k};\alpha^2\Bigr )
\\ \nonumber 
&& \qquad +\alpha^2 B \Bigl (\frac{3}{2},\frac{1}{k} \Bigr ) \phantom{a}_2F_1\Bigl (1+\frac{1}{k},\frac{3}{2},\frac{3}{2}+\frac{1}{k};\alpha^2\Bigr ) \>.
\ea
which, when substituted into Eq. \eqref{eq:Qa} is the equation we solve to obtain  $\omega $ in terms of $k$, $Q$, $m$ and~$g$.  Here $B(x,k)$ denotes the Beta function. 

In order to see if the classical solution describes a bound  state, one must 
calculate the value of the Hamiltonian for this solution and show that it 
is less than $m$.  The Hamiltonian density is given by \eqref{eq:hdensity}, so that the energy of the solitary wave is given by 
 \ba
 H_{sol} &=& \int  dx \ { \cal{H} } =   \int dx \  (h_1+h_2 -h_3) 
 \nonumber \\ 
 &=& \int dx \  \left [ h_1 \left(1-\frac{1}{k}\right) + h_2 \right ] 
 \nonumber \\ 
 &=& 
 H_1 \Bigl (1-\frac{1}{k} \Bigr ) +H_2 \>, 
 \ea
 where we have used Eq. \eqref{eq:relation}.  We find
 \ba
H_1 &=&  \int dy \  R^2(y) \frac{d\theta(y)}{dy} \nonumber \\
&=&\frac{\beta_{k}}{k(m+\omega )} \left [ \frac{(k+1)\beta_{k}^2}{g^2k^2(m+\omega )} \right ]^{\frac{1}{k}} 
\\ \nonumber
&& \times \ B \Bigl (\frac{1}{2},1+\frac{1}{k}\Bigr ) \phantom{a}_2F_1\Bigl (1+\frac{1}{k},\frac{1}{2},\frac{3}{2}+\frac{1}{k};\alpha^2\Bigr ) \>. \nonumber \\
H_2 &=& m\int dy \  R^2(y) \cos 2 \theta(y) \nonumber \\
&=&\frac{1}{\beta_{k}} \left [\frac{(k+1)\beta_{k}^2}{g^2k^2(m+\omega )} \right ]^{\frac{1}{k}} 
\\ \nonumber
&& \times \ B \Bigl (\frac{1}{2},\frac{1}{k}\Bigr ) \phantom{a}_2F_1\Bigl (\frac{1}{k},\frac{1}{2},\frac{1}{2}+\frac{1}{k};\alpha^2\Bigr ) \>. 
\ea
Without loss of generality, in the remaining part of this subsection we now 
put $m=1$ so that $0 \le \omega  \le 1$, in order to measure $\omega,H$ in units of $m$. For $k=1$ we find
\ba
&&H_1=\frac{2(1+\omega )}{g^2 Q}[(1+\alpha^2)\tanh^{-1}\alpha - \alpha] \>, \nonumber \\
&&H_2=\frac{4}{g^2 Q} \tanh^{-1} \alpha \>. \label{eq:hk1}
\ea
Therefore we find that the energy of the solitary wave is
\bq
E_{sol} = \frac{4}{g^2 Q} \tanh^{-1} \alpha_{sol}, 
\eq
where in $\alpha_{sol}$, $\omega_{sol} =  \frac{1}{\sqrt{1+Q^2 g^4/4}}$. 
We notice that the energy of the solitary wave with $k=1$ does not depend on the width parameter $\beta$. Simplifying we obtain for $k=1$ and
for all values of $g^2$
\bq
H_{sol} = \int dx \  \bpsi \psi = \frac{2}{g^2 Q}\sinh^{-1} (g^2 Q/2) < 1 ,
\eq
so that all the solutions are ``bound states".  This agrees with the result of Lee \emph{et al.} \cite{ref:Lee}.

\begin{figure}[t]
   \centering
   \includegraphics[width=3.4in]{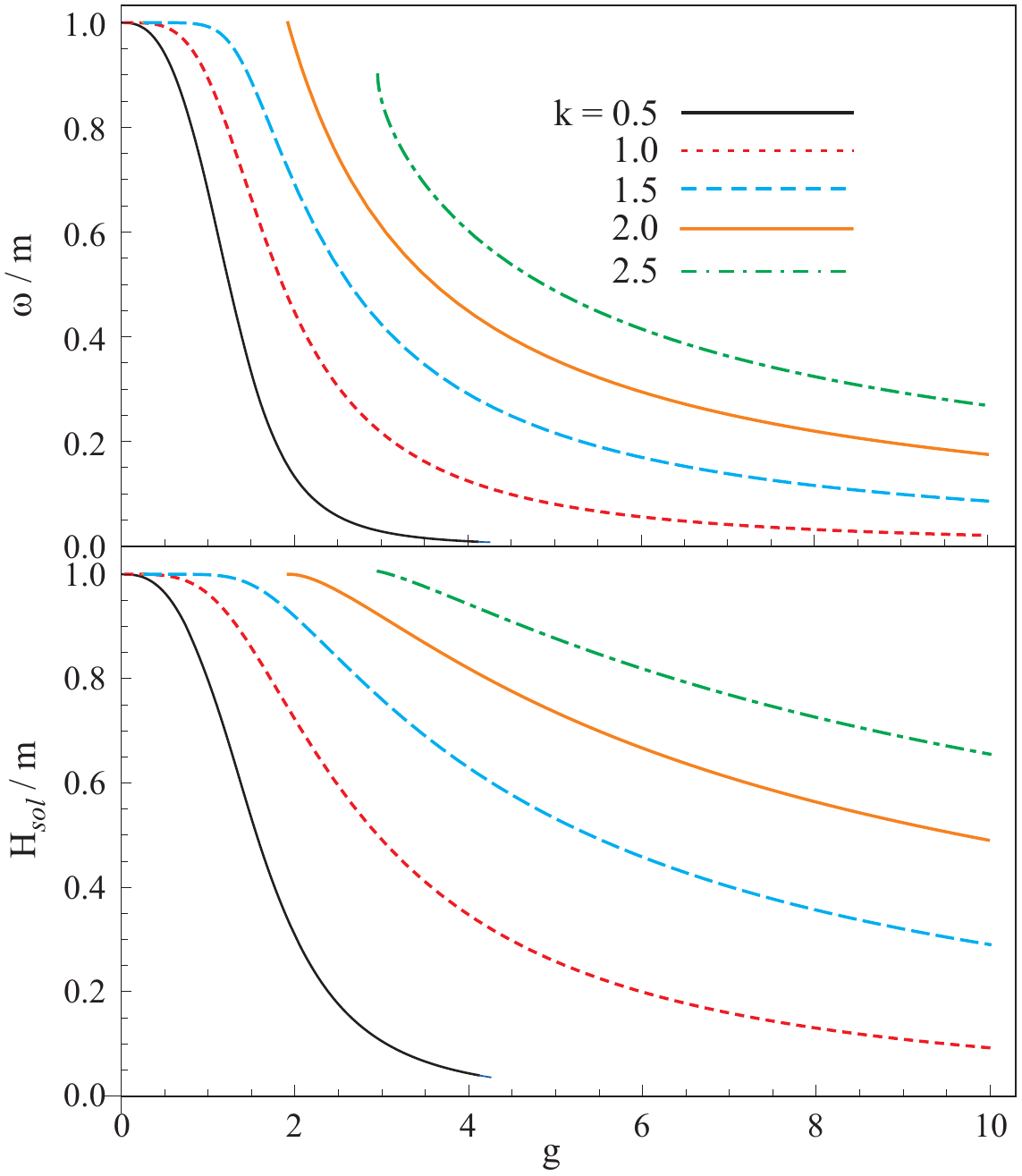}
   \caption{\label{dirac_ss}(Color online)
  NLDE bound states for the scalar-scalar interaction case: $\omega$ and 
$H_{sol}$ as a function of $k$ and~$g$ for $Q=1$.}
\end{figure}

\begin{figure}[t]
   \centering
   \includegraphics[width=3.4in]{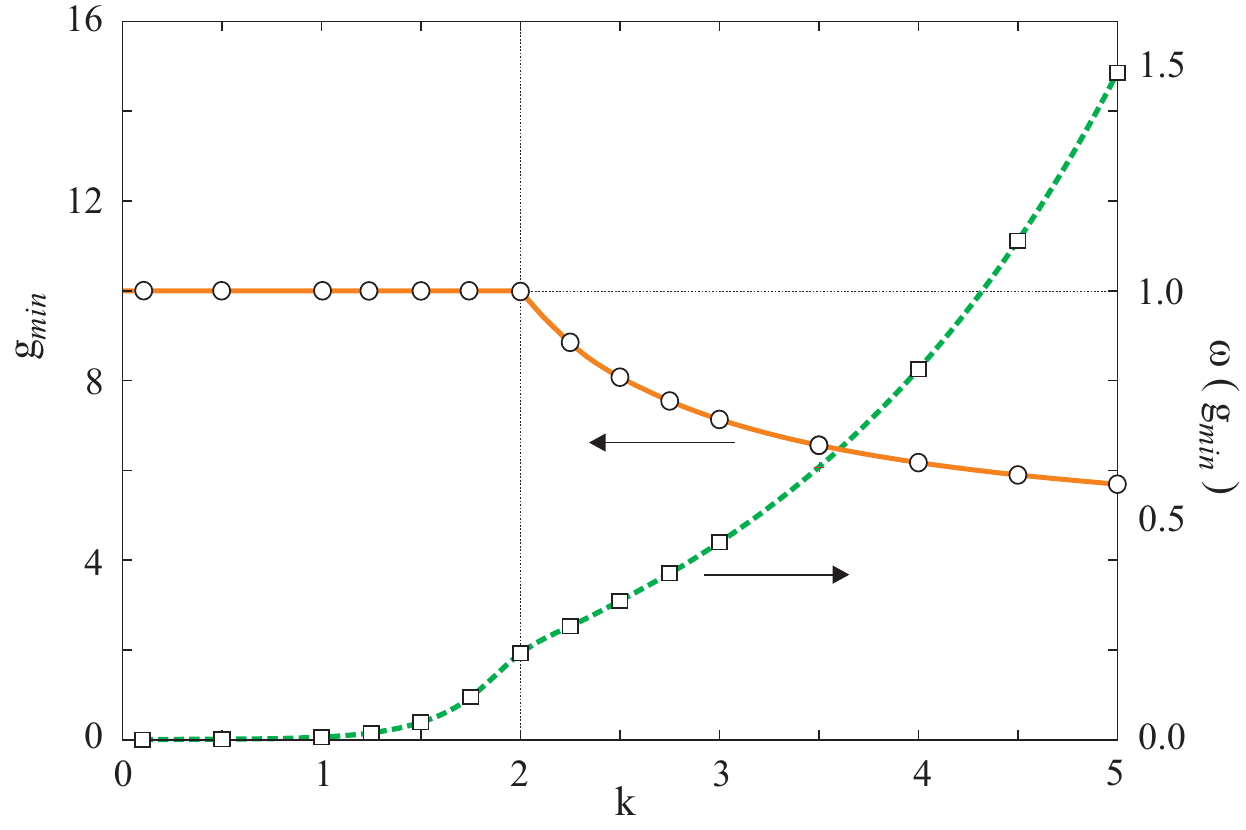}
   \caption{\label{dirac_ss_gs}(Color online)
   Plot of the lower bound of the allowed range of $g$ values in the scalar-scale interaction case, as a function of $k$, together with the corresponding solutions, $\omega(g_{min})$. The solid lines are intended only as a guide to the eye.}
\end{figure}

For $k=\frac{1}{2}$ one finds that
\begin{align}
H_1= & \frac{9(1-\omega)^2}{16g^4 Q}  [(3\alpha^4+2\alpha^2+3)\tanh^{-1}  \alpha  -3\alpha (1+\alpha^2) ] \>, \nonumber \\
H_2 = & \frac{9(1+\omega)}{2 g^4 Q } [- \alpha + (1+ \alpha^2) \tanh^{-1}  \alpha ] \>.  \label{eq:hhalf}
\end{align}
For $Q=1$ and  selected values of $k$ we determine $\omega $ and~$H_{sol}$ and 
plot in Fig.~\ref{dirac_ss} the allowed values for which $H_{sol} < 1$. Note
that the range of $g$ values for the existence of a bound state, as a function 
of $k$, is bounded from below. The functional dependence of the lower bound 
$g_{min}$, together with the corresponding solution $\omega(g_{min})$, as a 
function of $k$, are depicted in Fig.~\ref{dirac_ss_gs}. We note the rapid 
increase of $g_{min}$ at large values of $k$.  At $k \approx 2$, the upper 
bound of the solution $\omega(g_{min})$ becomes lower than 1, and we notice 
an inflection in $g_{min}(k)$. Summarizing, we find that in the S-S case, 
bound states exist for all values of $k$ and~$g>g_{min}$.

\subsection{Vector-Vector interaction}

For the V-V interaction case, we obtain
\bq
L_I =   \frac{ g^2}{k+1} ( {\bar \Psi} \gamma_\mu \Psi 
\bPsi \gamma^\mu \Psi)^{\frac{1}{2}(k+1)}=  \frac{g^2}{k+1} R^{2(k+1)} \>.
\eq
Eq. \eqref{eq:t11}  now becomes 
\bq
\omega R^2 - mR^2 \cos 2 \theta +  \frac{g^2}{k+1} R^{2(k+1)}=0 \>.
\eq
Thus
\bq
R^2 = \left[ \frac{(k+1) ( m \cos 2 \theta-\omega )}{g^2} \right]^{\frac{1}{k}} \>.
\eq 
This can be rewritten in the following two forms:
\ba
&&R^2 =  \left[ \frac{(k+1) \beta_{k}^2}{g^2 k^2 (\omega + m \cosh 2 \beta_{k} x)} \right]^{\frac{1}{k}} \nonumber \\
&&= 
 \left[ \frac{(k+1) \beta_{k}^2 {\rm sech}^2 \beta_{k} x }{g^2 k^2(m+\omega ) ( 1+\alpha^2 \tanh^2\beta_{k} x )} \right ]^{\frac{1}{k}}  \>. \label{eq:rhov}
\ea

The equation for $ \omega$   can then be determined by using the charge defined in Eq. \eqref{eq:Q}. This gives

\bq
Q=  \frac{1}{ \beta_{k} } \left [ \frac{(k+1) \beta_{k}^2}{g^2 k^2 (m+\omega )} \right ]^{1/(k)} \hat I_k[\alpha^2]  \>,
\eq
where
\bq
\hat I_k[\alpha^2] 
=
B \Bigl (\frac{1}{2},\frac{1}{k}\Bigr ) \phantom{a}_2F_1\Bigl (\frac{1}{2},\frac{1}{k},\frac{1}{2}+\frac{1}{k}; -\alpha^2\Bigr ) \>.
\eq
For $k=1$, this gives
\bq
Q=\int_{-\infty}^{\infty} dx \ R^2=\frac{4\tan^{-1}\alpha}{g^2} \>,
\quad
\omega =m\cos(g^2 Q/2) \>.
\eq
This imposes the restriction on the coupling constant, i.e. $g^2 Q < \pi$ so that the
spectrum is composed of positive-energy fermion states.  
On the other hand, for $k=\frac{1}{2}$ this gives
\bq
Q=\frac{9(1+\omega)}{2g^4}[\alpha-(1-\alpha^2)\tan \alpha ] \>.
\eq

\begin{figure}[t]
   \centering
   \includegraphics[width=3.4in]{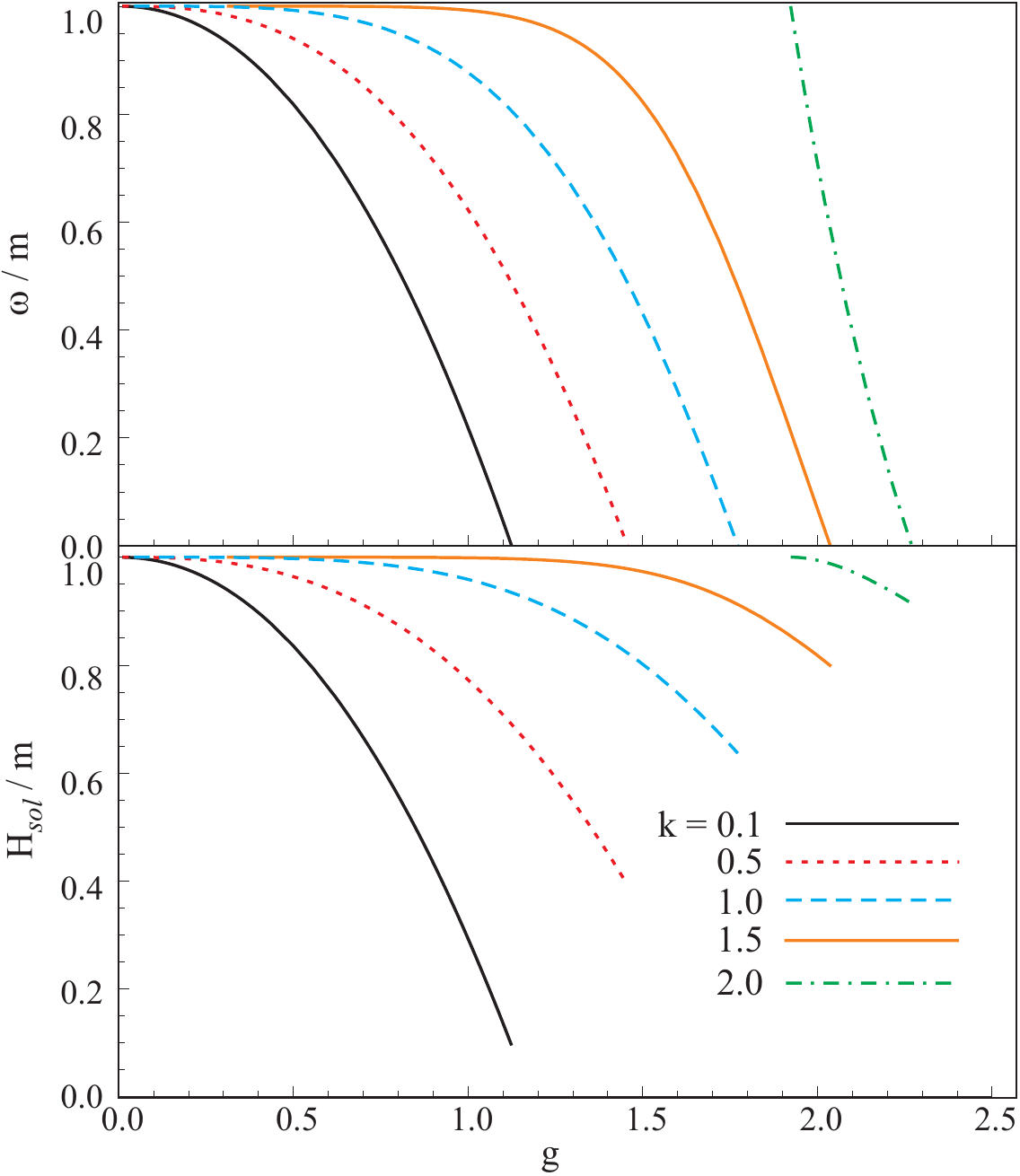}
   \caption{\label{dirac_vv}(Color online)
  NLDE bound states for the vector-vector interaction case: $\omega $ and 
$H_{sol}$ as a function of $k$ and~$g$.}
\end{figure}

\begin{figure}[t]
   \centering
   \includegraphics[width=3.4in]{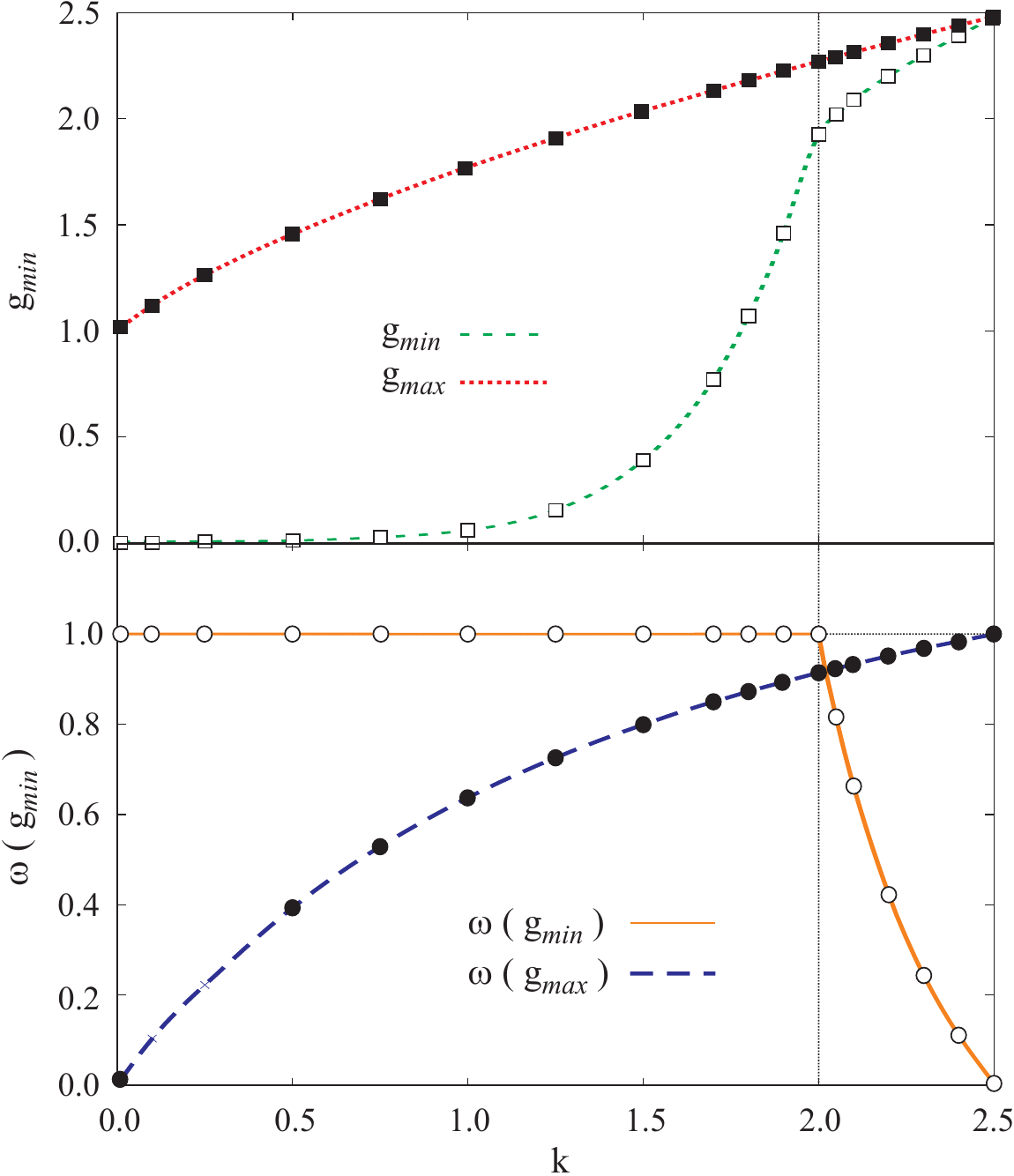}
   \caption{\label{dirac_vv_gs}(Color online)
   Plot of the lower and upper bounds of the allowed range of $g$ values in 
the vector-vector interaction case, as a function of $k$ (top panel). In the 
lower panel we depict the $k$ dependence of the corresponding solutions 
$\omega(g_{min})$ and~$\omega(g_{max})$, respectively. The solid lines are 
intended only as a guide to the eye.}
\end{figure}

The energy of the solitary wave is given by integrating the Hamiltonian density \eqref{eq:hdensity}, and we obtain
 \ba
 H_{sol} &=& \int  dx \  {\cal{H}}= \int dx \  (h_1+h_2 -h_3) 
 \nonumber \\
 &=& H_1 \Bigl (1-\frac{1}{k} \Bigr )  +H_2 \>,
 \ea
where
\ba
H_1&=&\frac{\beta_{k}}{k(m+\omega )}\left [ \frac{(k+1)\beta_{k}^2}{g^2k^2(m+\omega )} \right ]^{\frac{1}{k}} 
\\ && \times \ B \Bigl (\frac{1}{2},1+\frac{1}{k}\Bigr ) \phantom{a}_2F_1\left (1+\frac{1}{k},\frac{1}{2},\frac{3}{2}+\frac{1}{k}; -\alpha^2\right) \>, \nonumber \\
H_2&=&\frac{1}{\beta_{k}}\left [ \frac{(k+1)\beta_{k}^2}{g^2k^2(m+\omega )} \right ]^{\frac{1}{k}} 
\\ \nonumber && \times \ B \Bigl (\frac{1}{2},\frac{1}{k}\Bigr ) 
\Bigl [ 2 \  _2F_1\left (\frac{1}{k},\frac{1}{2},\frac{1}{2}+\frac{1}{k}; -\alpha^2\right ) 
\\ \nonumber && 
\qquad \qquad \qquad - \phantom{a}_2F_1\Bigl (\frac{1}{2},\frac{1}{k},\frac{1}{k}+\frac{1}{2}; -\alpha^2\Bigr ) \Bigr ] \>.
\ea
Without any loss of generality, in the remaining part of this subsection we
put $Q=m=1$, i.e. we measure $\omega,H$ in units of $m$ so that 
$0 \le \omega \le 1$. For $k=1$, we find 
\ba
H_1&=& \frac{2(1+\omega)}{g^2}[\alpha-(1-\alpha^2) \tan^{-1} \alpha] \>,
\\ \nonumber 
H_2 &=& \frac{4\alpha}{g^2(1+\alpha^2)} \>.
\ea 
For $k=1$, we have an analytic solution: 
\ba
H_{sol} =  \int dx \ \bpsi \psi  = \frac{2}{g^2}\sin (g^2/2) < 1 \>,
\ea
since $0 < g^2 < \pi$, thereby showing the bound-state behaviour even in the 
vector case.  
For $k=\frac{1}{2}$, one finds
\begin{align}
H_1 =& \frac{9(1+\omega)^2}{16g^4}[(3\alpha^4-2\alpha^2+3)\tan^{-1} \alpha -3\alpha(1-\alpha^2)] \>,
\nonumber \\
H_2 =& \frac{9(1+\omega)}{4g^4}[(1+\alpha^2)\tan^{-1}  \alpha -\frac{\alpha (1-\alpha^2)}{(1+\alpha^2)}] \>,
\nonumber \\
H_3 =& 2H_1 \>.
\end{align}
In Fig.~\ref{dirac_vv} we map out the allowed values of $\omega$  and~$g^2$  
for various values of $k$. The allowed range of $g$ values for the existence 
of a bound state, as a function of $k$, has both a lower and an upper bound, 
and the domain shrinks as~$k$ increases. Around $k$=2.5, these bounds cross, 
and no bound states are possible for $k > 2.5$. The functional dependence of 
$g_{min}$ and~$g_{max}$, together with the corresponding solutions 
$\omega(g_{min})$ and~$\omega(g_{max})$, as a function of $k$, are depicted in 
Fig.~\ref{dirac_vv_gs}. As in the S-S case, $\omega(g_{min})$ becomes less 
than 1 for $k \approx 2$, and we notice an inflection in $g_{min}(k)$. However,
we now find that $\omega(g_{max})$ approaches one in case $k>2$. 

\section{Connection to the solutions of the NLSE}

In this section we will perform the nonrelativistic reduction of the NLDE to determine how it compares to the NLSE. The NLDE can be written as
\bq
i  \sigma_3  \partial_t \Psi + \sigma_x \partial_x \Psi - m \Psi - V_I \Psi =0 \>.  \label{NLDEb} ,
\eq
where $ V_I = - \frac{\partial L_I}{\partial {\bar \Psi} } = -g^2 (\bpsi \psi)^k$. 
Next, we use Moore's decoupling method \cite{ref:Moore} and write
\bq
V_I[\lambda] = \frac{1+ \sigma_3}{2 } V_I + \lambda  \frac{1-\sigma_3}{2 } V_I \>.
\eq
We see that $V_I[ \lambda=1] = V_I$.  It has been shown that doing a 
perturbation theory in $\lambda $ is  a valid way of obtaining the corrections 
to the nonrelativistic theory. Moore's decoupling technique was used for the 
(relativistic) hydrogen atom using conventional Rayleigh-Schr\"odinger 
perturbation theory and computer algebra and it was shown that the perturbative
solution converges to the correct solution \cite{ref:Moore}. It has been 
applied successfully to the relativistic calculations on alkali atoms and 
represents one of the many relativistic perturbative schemes investigated by  
Kutzelnigg \cite{ref:Kutz}.  We will show that this procedure  leads to the 
heuristically derived nonrelativistic reduction of the NLDE  as discussed by 
Toyama \emph{et al.} for the case $k=1$ \cite{ref:Toyama}. 

We  let
\ba
\Psi_0 (x) &&=   e^{-i\omega t}  \left(\begin{array}{c}u_0 \\v_0\end{array}\right)  \>,
\ea
be a solution of the theory when $\lambda = 0$. 
For scalar-scalar interactions, we find: 
\ba
&& \frac{du_0}{dx} + (m+\omega ) v_0 =0 \>, \nonumber \\  \label{nr}
&&\frac{dv_0}{dx} + (m-\omega ) u_0 - g^2(u_{0}^\star u_0-v_{0}^\star v_0)^{k}
 u_0=0 \>. \nonumber \\
\ea
From Eq.~\eqref{nr} we obtain
\bq
 \frac{du_0}{dx} = - (m+\omega ) v_0 \>.
\eq
This leads to the following equation for $u_0$:
\bq
  -  \frac{(u_0)_{xx}}{2m} + (V_I-\epsilon_0)\left( 1+ \frac{\epsilon_0}{2 m}\right) u_0=0 \>,
\eq
where $ \epsilon_0= \omega - m$.  We notice that the expansion parameter 
is $\epsilon_0/(2m)$.  When  $|\omega - m|/(2m)\ll 1$
 is satisfied then we can be sure that the NLDE solutions go over to the 
NLSE solutions.  However we will find that in the V-V case, the reduction 
numerically appears valid over a wider range.  
The relevant Schr\"odinger-like equation is:
\bq
  -  \frac{(u_0)_{xx}}{2m} +  \hat {V}_I u_0 = {\hat E} u_0 \>,
\eq
where
\bq
  \hat {V}_I = V_I  \Bigl ( 1+ \frac{\epsilon_0}{2 m} \Bigr ) \>, \qquad 
  {\hat E} = \epsilon_0  (1+ \frac{\epsilon_0}{2 m})
  \>.
\eq
For consistency we need to expand~$V_I$ to first order in $1/2m$.  For the 
scalar scalar case, we have
\ba
V_{I} &&= - g^2 (u_{0}^\star u_0-v_{0}^\star v_0)^{k}  \nonumber \\
&& \rightarrow  -g^2   \left[ (u_{0}^\star u_0)^{k} - \frac{k}{4 m^2} (u_{0}^\star u_0)^{k-1} (u_0)_x^\star (u_0)_x \right] . \nonumber\\
\ea
The resulting modified nonlinear Schr\"odinger equation (mNLSE) can be derived 
from the Lagrangian:
\ba
L = && i \psi^\star \partial_t \psi - \frac{1}{2m} \left[ \psi^\star_x \psi_x\left(1+ \frac{{\hat g}^2}{2m} (\psi^\star \psi)^k \right) \right] \nonumber \\
&& + \frac{{\hat g}^2}{k+1} (\psi^\star \psi)^{k+1}, 
\ea
and the Hamiltonian is given by 
\bq
H = \!\! \int  \frac{dx}{2m} \left[ \psi^\star_x \psi_x\left(1+ \frac{{\hat g}^ 2}{2m} (\psi^\star \psi)^k \right) \right] 
-\frac{{\hat g}^2}{k+1} (\psi^\star \psi)^{k+1} \>,
\eq
where 
$ {\hat g}^2 = g^2 [1+ \epsilon_0/(2m)]$. 

In the case of the V-V interaction, the nonrelativistic reduction of the NLDE is similar to the previous case with the difference that 
\ba
V_{I}^{v-v}  &&= - g^2 (u_{0}^\star u_0+v_{0}^\star v_0)^{k}  \nonumber \\
&& \rightarrow  -g^2   \left[ (u_{0}^\star u_0)^{k} + \frac{k}{4 m^2} (u_{0}^\star u_0)^{k-1} (u_0)_x^\star (u_0)_x \right] . \nonumber \\
\ea
The resulting modified nonlinear Schr\"odinger equation (mNLSE) can be derived from the Lagrangian:
\ba
L = && i \psi^\star \partial_t \psi - \frac{1}{2m} \left[ \psi^\star_x \psi_x\left(1- \frac{{\hat g}^2}{2m} (\psi^\star \psi)^k \right) \right] \nonumber \\
&& + \frac{{\hat g}^2}{k+1} (\psi^\star \psi)^{k+1},
\ea
and the Hamiltonian is given by 
\bq
H = \!\! \int \frac{dx}{2m} \left[ \psi^\star_x \psi_x\left(1- \frac{{\hat g}^ 2}{2m} (\psi^\star \psi)^k \right) \right] 
-\frac{{\hat g}^2}{k+1} (\psi^\star \psi)^{k+1}, 
\eq
where 
$ {\hat g}^2 = g^2 [1+ \epsilon_0/(2m)]$.
 
Thus we see that the resulting theory in the large $1/2m$ limit  (as well 
as when $|\omega - m| \ll 2m$), in both S-S and V-V cases reduces to the modified 
NLSE equation. The first correction has the same magnitude but opposite sign for 
the two cases.   

\subsection{ Comparison with the exact solution of the NLSE and mNLSE}
Here we want to compare the NLDE with the exact solution of the of the NLSE 
as well as mNLSE for arbitrary~$k$. 
We will give numerical comparison both when the criterion $|\omega - m| \ll 2m$ 
is satisfied and
for general $\omega$.  We will find that the V-V  NLDE case has solutions that 
track those of the NLSE for a broader range of $\omega$.

First let us obtain solutions to the NLSE  for arbitrary~$k$.  
The NLSE is defined by the Lagrangian 
\bq
L = {\frac{i}{2}} \int d x \ (\pstar \psi_{t} -\pstar_{t}\psi) - H  \label{eq:action}
\>,
\eq
where for the S-S interaction 
\bq
H = \int d x \   \left [  \frac{1} {2m} \nabla\pstar \nabla\psi -
g^2 {\frac{(\pstar\psi)^{k+1}}  {k+1}} \right ] \>.
\eq
This leads to the equation of motion
 \bq  \label{3.1} 
i \, {\frac{\partial \psi}{\partial t}} +   \frac{1} {2m} \left( \frac {\partial \psi }{\partial x} \right)^{2} +
 g^2 (\pstar\psi)^{k} \psi = 0 \>.
\eq
If we make the ansatz 
\bq  \label{3.2}
\psi(x,t)=r(y)\exp[i(m vy-\omega t+\delta)]\,,~~y=x-vt\,,
\eq
then it is easy to show that $r(y)$ satisfies the equation
\bq  \label{3.3}
r''(y) - \Omega r(y)+g^2r^{2k+1}(y)=0\,,
\eq
where $\Omega=-(\omega+\frac{mv^2}{2})$. 
 Equation (\ref{3.3}) has an exact solution
\bq  \label{3.4}
r(y)=A\sech^{1/k}[D(y+y_0)]\,,
\eq
provided
\bq  \label{3.5}
\Omega= \frac{D^2}{2m k^2 }\>, \quad A^{2k}=\frac{(k+1)D^2}{2m g^2  k^2}\,.
\eq
The mass density in the rest frame ($v=0$)  is given by 
\bq\label{3.6}
\rho = \pstar  \psi = \left[ \frac{(k+1)D^2 }{2m g^2  k^2}
                                  \right]^{1/k} \sech^{2/ k}[  D  (x+x_0)] \>. 
\eq

Let us now obtain the solutions of the mNLSE. We first notice that to the first
order in $1/2m$, the static mNLSE equation in both S-S and V-V cases is 
given by
\bq\label{3.7}
-(u_0)_{xx}+(m^2-\omega^2) u_0-(m+\omega)g^2[u_{0}^{\star}u_0]^{k}u_0=0\,,
\eq
which has the exact solution
\bq\label{3.8}
u_0(x)=A\sech^{1/k}[\beta_k(x+x_0)]\,,
\eq
with
\bq  \label{3.9}
A^{2k}=\frac{(k+1)\beta_k^2}{(m+\omega) g^2  k^2}\,.
\eq
Hence for mNLSE, the  mass density in the rest frame ($v=0$)  is given by 
\bq\label{3.10}
\rho = \pstar  \psi = \left[ \frac{(k+1)\beta_k^2 }{(m+\omega) g^2  k^2}
\right]^{1/k} \sech^{2/ k}[\beta_k (x+x_0)] \>. 
\eq

We will now compare the NLSE and mNLSE solutions with the solutions of the 
NLDE. In making these comparisons we will in all cases compare the solutions 
for the charge density (which is the mass density for the NLSE and mNLSE).

\subsection{Scalar-Scalar interaction}

One can rewrite the charge density $\rho = R^2$, Eq. ~\eqref{eq:Rsq} in the following form which isolates the previous solution to the NLSE. 
\ba
&&\rho = \left[ \frac{\beta_k^2 (k+1)}{g^2 k^2(m+\omega)} \right]^{1/k} {\rm sech} ^{2/k} \beta_k x~~ f(\alpha, \beta,x),  \nonumber \\
&&f(\alpha, \beta,x)= \frac  {1+\alpha^2 \tanh^2\beta_{k} x }  {(1-\alpha^2 \tanh^2\beta_{k} x )^{(1+1/k)} }
  \>.
  \label{eq:Rsqa}
\ea
If we compare NLSE and S-S case, we find that $\rho(x=0)$ is same in both cases
only if we can identify $D$ with $\beta_k$. We also have that  
$f(\alpha, \beta,x=0) =1$, so that with this identification, the charge and 
mass densities have the same value as a function of $k$ for the NLSE 
and NLDE.   

On the other hand, $\rho(x=0)$ is strictly identical for S-S and mNLSE cases
and no identification needs to be made.

\begin{figure}[t]
   \centering
   \includegraphics[width=3.4in] {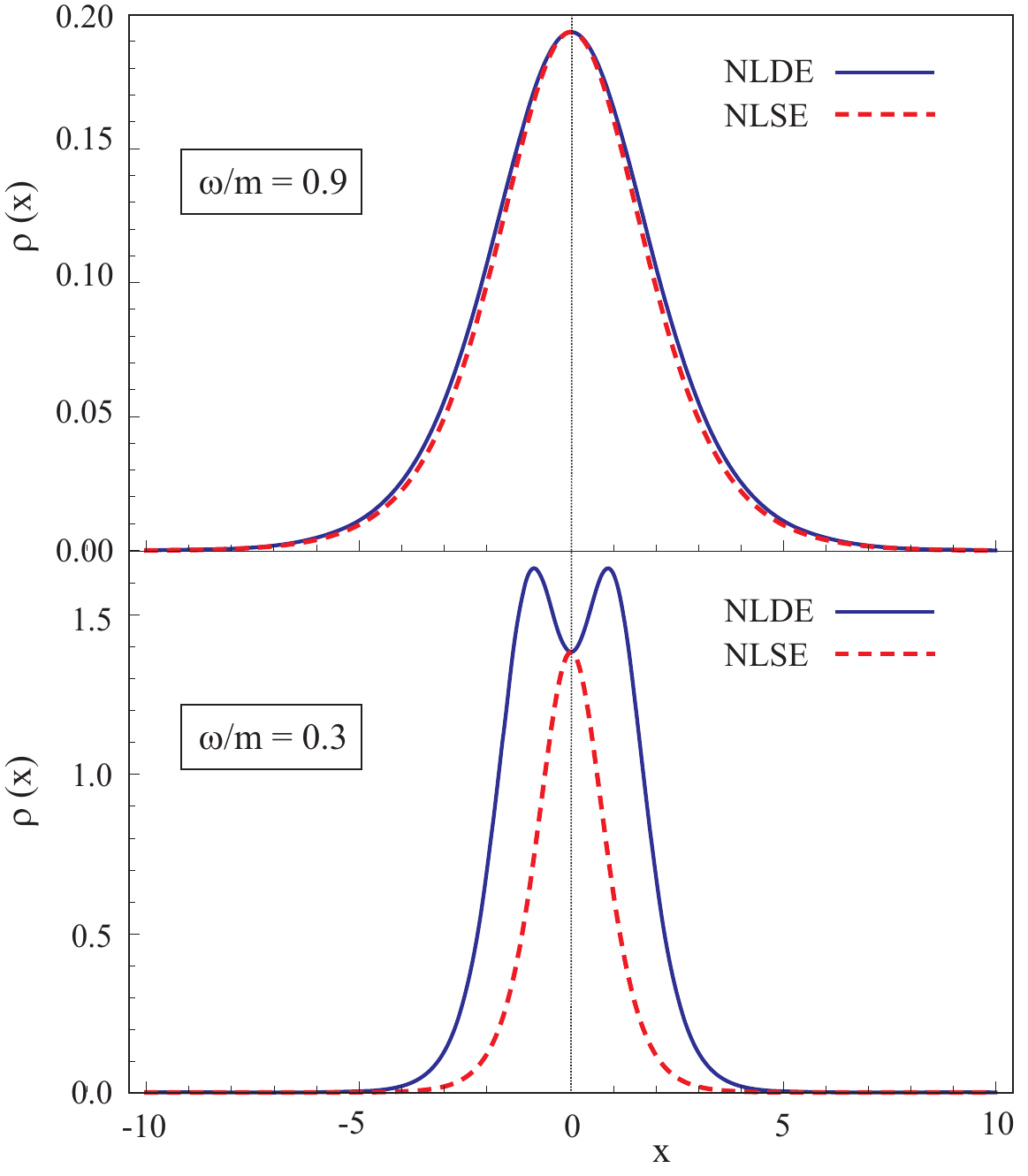}
   \caption{\label{scalar_nlse} (Color online)
 Comparison of the NLSE and NLDE solutions in the case of scalar-scalar 
interactions for  $k=1$,  and  $\omega/m = 0.9$ (top panel) and 
$\omega/m = 0.3$ (bottom panel), respectively.}
\end{figure}

\begin{figure}[t]
   \centering  
   \includegraphics[width=3.4in] {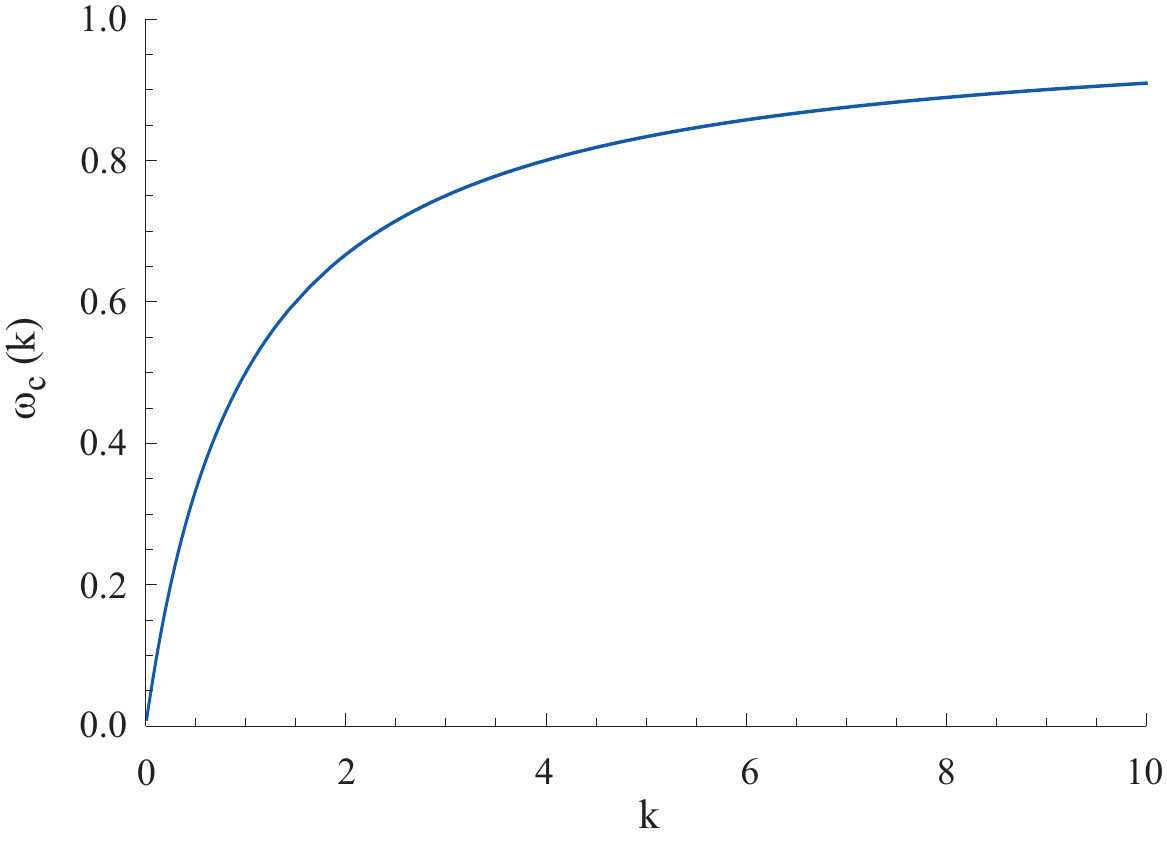}
   \caption{\label{omegac}(Color online)
 Critical value, $\omega_c(k)$, for any $\omega \le \omega_c(k)$ the solution 
of the NLDE equation becomes double humped in the case of scalar-scalar 
interactions.}
\end{figure}

We have seen that the nonrelativistic limit is obtained when 
$|\omega -m| /2m \ll 1$.  In Fig.~\ref{scalar_nlse},  we compare the solutions 
to the NLSE and NLDE when $\omega/m = 0.9$ (top panel) and~$\omega/m = 0.3$ 
(bottom panel), for $k=1$. In the latter case, we notice that the solution to 
the NLDE is double humped.  For any $ \omega \le \omega_c(k)$ 
for which the 
solution becomes double humped in the NLDE is shown in Fig.~\ref{omegac}.

\subsection{Vector-Vector interaction}

Now we  rewrite the solution found for the charge density $\rho = R^2$, Eq.~\eqref{eq:rhov} in the following form:
\ba
&&\rho = \left[ \frac{\beta_k^2 (k+1)}{g^2 k^2(m+\omega)} \right]^{1/k} {\rm sech} ^{2/k} \beta_k x~~ f(\alpha, \beta,x) \>,
\nonumber \\
&&f(\alpha, \beta,x)=\  {(1+\alpha^2 \tanh^2\beta_{k} x )^{-1/k} }
  \>.
  \label{eq:Rsqb}
\ea

\begin{figure}[t]
   \centering
   \includegraphics[width=3.4in] {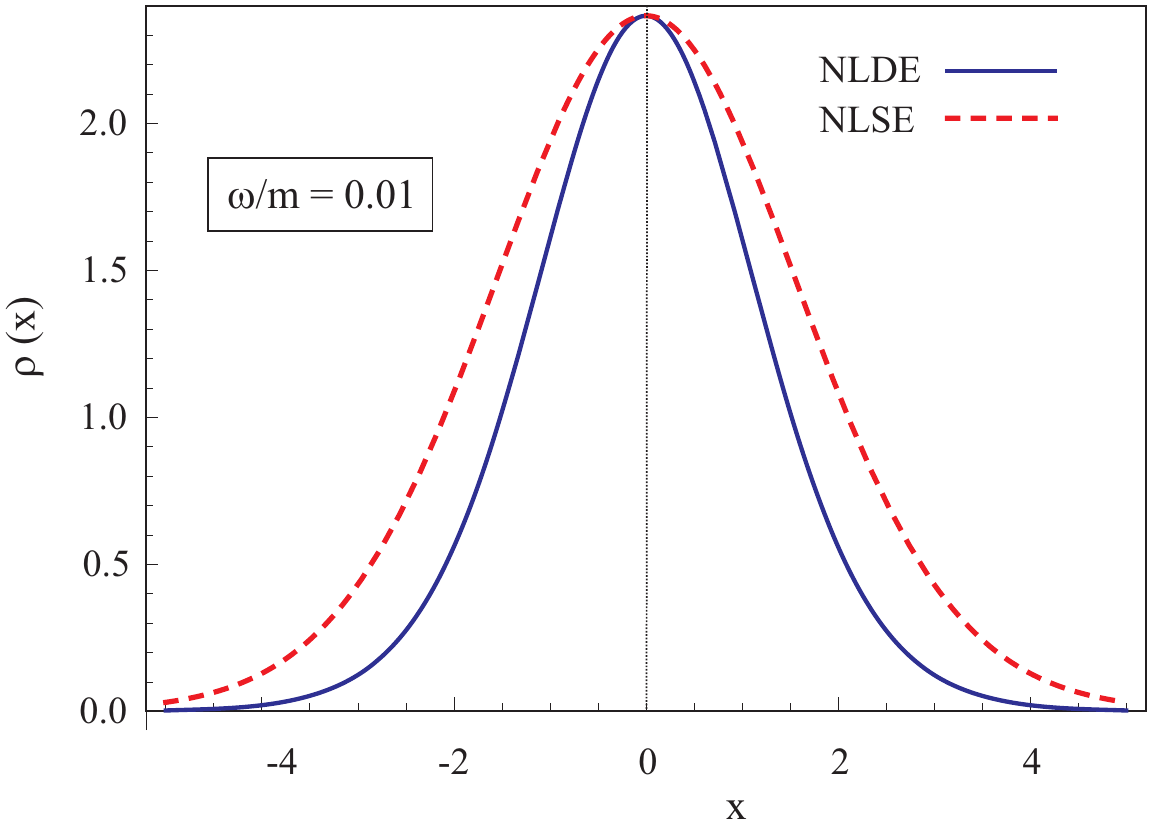}
   \caption{ \label{vector_nlse}(Color online)
  Comparison of the NLSE and NLDE solutions in the case of vector-vector interactions for  $k=1$ and  $\omega/m = 0.01$.}
\end{figure}

We have seen that the nonrelativistic limit is obtained when 
$|\omega -m| /2m \ll 1$.
For the V-V case, the modification of the NLSE result is small even at very 
small $\omega/m$ and, unlike in the case of S-S interactions, the NLDE 
solution never becomes double humped.
In Fig.~\ref{vector_nlse}  we compare the solutions of the NLSE and NLDE  
when $\omega/m=0 .01$  and~$k=1$. The main difference compared to the S-S
case is that the convergence to the nonrelativistic limit 
as $\omega/m \rightarrow 1$, occurs from above in the vector case instead 
of from below as in the scalar case (see Fig.~\ref{vector_nlse} and the top 
panel of Fig.~\ref{scalar_nlse}).
Again notice that $\rho(x=0)$ is identical in NLSE and V-V case only if we
identify $D$ with $\beta_k$. 

On the other hand, $\rho(x=0)$ is strictly identical for  mNLSE and V-V case 
and no identification needs to be made.

\section{stability of static solutions}
The stability of the solitary waves of the NLSE have been studied for a long 
time.  A recent discussion of this is found in \cite{review}.  
In this section we will first show that an analysis of the solutions of the 
NLSE equation using the slope criterion ($ \frac{ dM(\omega)}{d \omega} < 0$ 
for stability) where $M$ is the mass of the Solitary wave and   $\omega$  
the frequency gives the same result ($0<k<2$)  as  an analysis   based on 
whether a  scale transformations  raises or lowers the energy of the solitary 
wave.  The latter criterion is similar to the arguments first used by 
Derrick \cite{ref:derrick} in his study of the relativistic scalar field 
theories. We will then use a similar scaling argument first made by  Bogolubsky
 \cite{ref:bogol} for the NLDE equation to obtain a criterion for stability. We
 will find that the results of this approach do not agree with a smooth 
continuation of the result for the NLSE. We will discuss the most likely 
reason for the failure of this method when applied to the NLDE. 
Finally we will look at the stability in the mNLSE which contains the first 
relativistic correction to the NLSE and show that it gives essentially the 
same criterion as that found for the NLSE, i.e.  when $0<k<2$ we expect 
the solutions to be stable. 

Most studies of the stability of static solutions of the NLSE rely on the 
existence of a variational principle
\bq
\delta {\mathcal E}  = \delta(H- \omega M) =0 \>, \label{var1a}
\eq
from which the ordinary differential equation for  the  solution  
$u(x,\omega)$ can be derived. Here
 the NLSE Hamiltonian is 
 \bq
 H = \int dx \  \left[ \frac{1}{2m} \partial_x \psi^\star \partial_x \psi - \frac{g}{k+1}  (\psi^\star \psi)^{k+1} \right]  \>,
 \eq
 and the mass is given by 
 \bq
 M = \int dx \psi^\star \psi \>.
 \eq
 
This variational principle is quite similar to the one used to study the 
stability in the generalized KdV systems \cite{K,DK,VK,cooper3}. There 
one derives the solitary wave equation from
\bq
\delta\epsilon=\delta(H-cP)=0\,,
\eq
where $c$ is the velocity of the solitary wave while the generalized KdV
equation is
\ba
&&u_t+u^{l-2}u_x \nonumber \\
&&+\alpha[2u^{p}u_{xxx}+4pu^{p-1}u_xu_{xx}+p(p-1)u^{p-2}
(u_x)^{3}]=0\,. \nonumber \\
\ea
This can be derived from the Hamiltonian
\bq
H=\int dx \left[-\frac{u^l}{l(l-1)}-\alpha u^p (u_x)^2\right]\,,
\eq
and the corresponding momentum P is given by
\bq
P=\int dx \frac{1}{2}u^2(x,t)\,.
\eq

Stable solitary waves of the form $\psi(x,t) = u(x, \omega) e^{-i \omega t} $ 
need to be local energy minimizers of the  functional \eqref{var1a}. 
Based on linearized perturbation theory  and using this variational principle 
Vakhitov and Kolokolov \cite{VK} showed  that a necessary 
 criterion for stability is that 
 \bq
 \frac{ dM(\omega)}{d \omega} < 0.
 \eq
This criteria is the analogue of  the result found for the generalized  KdV 
equations by  Karpman \cite{K}  and Dey and Khare \cite{DK} who obtained 
that stable solitary waves for that system of equations required
 \bq
 \frac{ dP(c)}{d c} >0.
 \eq
 
The exact solution for $u(x)$ of NLSE for arbitrary~$k$ is  given in  
\eqref{3.4}.  Using that solution, one finds that  the mass  has the 
following dependence on $\omega$:
 \bq 
M = C (-\omega)^{(2-k)/(2k)} \>,
\eq
and the necessary criterion for stability  is
\bq
k < 2. \label{eq:k} \> 
\eq

Another approach to stability, which leads to the same result  
as \eqref{eq:k},  is based on whether a scale transformation which keeps the
mass $M$ invariant, raises or lowers the energy of a solitary wave. For the  
NLSE  with Hamiltonian 
\ba
H &=& \int dx \  \left[ \partial_x \psi^\star \partial_x \psi - \frac{g}{k+1}  (\psi^\star \psi)^{k+1} \right] 
\nonumber \\
&\equiv& H_1-H_2 \>,
\ea
both $H_1$ and~$H_2$ are positive definite. 
A static  solitary wave solution can be written as
\bq
\psi(x,t) = r(x) e^{-i\omega  t} \>.
\eq 

The exact solution has the property that it minimizes the Hamiltonian subject 
to the constraint of fixed mass as a function of a stretching factor  $\beta$.
This can be seen by studying a variational approach as done in \cite{cooper3} 
or by directly studying the effect of a 
scale transformation that respects conservation of mass. 

In the latter approach, which generalizes the method used by 
Derrick \cite{ref:derrick}, we let
\bq  x \rightarrow   \beta x \>,
\eq
and consider  
\bq 
\psi_\beta(x) = \beta^{\frac{1}{2}} r(\beta x)  e^{-i \omega t} \>,
\eq
this leaves
\bq
   M= \int dx \  \psi^\star \psi =  \int dx \  \psi_\beta^\star \psi_\beta \>,
 \eq
 unchanged.  One defines $H_\beta$ as the value of $H$ for the stretched 
solution $\psi_\beta$. One then finds that 
 \bq
  \frac{ \partial H_\beta}{\partial \beta} \Bigr |_{\beta=1} = 0 \>,
 \eq
 is consistent with the equations of motion, and the stable solutions 
satisfy 
 \bq
\frac{\partial^2  H_\beta } {\partial \beta^2}  \geq 0 \>.
\eq
If we write $H$ in terms of the two positive definite pieces $H_1$, $H_2$, then
\bq
H_\beta  = \beta^2  H_1-  \beta^{k} H_2 \>.
\eq
We find: 
\bq
\frac{\partial H_\beta} {\partial \beta}= 2 \beta H_1- (k) \beta^{k-1} H_2 \>.
\eq
We obtain
\bq
\frac{\partial H_\beta} {\partial \beta}|_{\beta=1 }  =  0 \rightarrow   H_1 = \frac{k}{2}  H_2 \>.
\eq
This result is consistent with the equation of motion.
The second derivative is given by 
\bq
\frac{\partial^2  H_\beta} {\partial \beta^2} = 2 H_1 - k(k-1) \beta^{k-2} H_2 \>,
\eq
which when evaluated at the stationary point  yields
\bq
\frac{\partial^2  H_\beta} {\partial \beta^2}= 2 (2-k)  H_1 \>.
\eq
This result indicates that solutions are unstable to changes in the 
width (compatible with the conserved mass) when $k > 2$.
The case $k=2$ is the marginal case where it is known that blowup occurs at 
a critical mass (see for example Ref.~\onlinecite{cooper3}). 
The result found above for the NLSE has also been found by various other 
methods such as linear stability analysis and using
strict inequalities. Numerical simulations have been done for the critical 
case  $k=2$ showing that blowup (self-focusing)  occurs when the mass 
$M > 2.72$~\cite{numerical}.  For $k>2$ a variety of analytic and numerical 
methods have been used to study the nature of the blowup at finite 
time~\cite{kevrekedis}. 

Let us now apply this scaling argument, as was done by 
Bogolubsky \cite{ref:bogol}), to the 1+1 dimensional NLDE. Again we will 
assume that  the exact solution minimizes   $H_\beta$ when $\beta =1$  with 
the constraint that 
the charge is kept fixed. (The validity of this assumption will be challenged 
below. All that is known is that $H_\beta$ is a stationary point at the 
solution.)

Our exact solution is of the form 
\bq
\psi(x) =  \left(\begin{array}{c}u \\v\end{array}\right)  = R(x) \left(\begin{array}{c}\cos \theta \\ \sin \theta \end{array}\right) e^{-i\omega t} \>.
\eq
Because we want to keep the charge fixed, we consider the following stretched 
solution:
\bq
\psi_\beta(x) =  \left(\begin{array}{c}u \\v\end{array}\right)  =  \beta^{\frac{1}{2}}  R(\beta x) \left(\begin{array}{c}\cos \theta(\beta x) \\ \sin \theta(\beta x)  \end{array}\right) e^{-i\omega t} \>.
\eq
The value of the Hamiltonian
\ba
H &=& \int dx \  \Bigl [ \bpsi i \gamma_1 \partial_1 \psi + m \bpsi \psi  - \frac {g^2}{k+1} (\bpsi \psi)^{k+1}  \Bigr ]
\nonumber \\
&\equiv& H_1 +H_2 - H_3 \>,
\ea
for the  stretched solution is 
\bq
H_\beta = \beta H_1 + H_2 - \beta^k H_3 \>,
\eq
where again $H_i$ are all positive definite. 
The first derivative is 
\bq
\frac{\partial H_\beta}{\partial \beta} = H_1- k \beta^{k-1} H_3 \>.
\eq
At the minimum,  setting $\beta =1$, we find in general 
\bq
H_3 = \frac{1}{k} H_1 \>,
\eq
which is consistent with the equation of motion result we obtained earlier, 
see Eq. ~\eqref{eq:relation}. We see that for $k=1$ the energy is given by 
just  $H_2$. The second derivative yields:
\bq
\frac{\partial^2 H_\beta}{\partial \beta^2} = - k(k-1) \beta^{k-2} H_3 \>. \label{second}
\eq
From this we see that if $k>1$, this analysis (if correct) would suggest that 
solitary waves are unstable to small  changes in the width.  For $0<k< 1$ the 
solitary waves are stable to this type of perturbation. 
This argument does not depend on $L_I$ as long as $L_I$ is positive definite. 
The same result is valid for both scalar and vector type interactions. 

For $k=1$, this argument does not give any insight into whether the solutions 
are stable.  However, it is  known that the solitary waves  discussed here 
for $k=1$, do appear to be stable numerically. Further, when they are 
scattered in numerical experiments, they interchange charge and energy, and 
sometimes show bound state production.  Detailed numerical simulations have 
been performed by  Alvarez and Carreras \cite{ref:numerical}.   These results
contradict the work of Bogolubsky \cite{ref:bogol} who studied changes in the 
frequency $\omega$ while keeping the charge fixed.  There a similar analysis gave 
a maximum for the Hamiltonian when $\omega < 1/\sqrt{2}$, even though, as
remarked above,  numerical studies show that solitary waves in that frequency 
range are in fact stable. 

We have already shown above that the solutions of the NLDE reduce to those of 
the NLSE in the nonrelativistic limit.  Assuming continuity arguments apply, 
one would expect that there would be at least a range of values of $\omega$ 
for which the solutions to the NLDE are stable for $k<2$.   

So one needs to understand the reason for this apparent discrepancy.  The main 
reason for assuming instability when
the second derivative  of $H(\beta)$  is positive, is  that the stable  
solutions to the Dirac equation are at least relative minima of the effective 
action. However, the study by Blanchard et al.  \cite{Blanchard} to find 
an analytic criterion for  stability in the 1+1 dimensional NLDE using the 
Shatah-Strauss formalism found that bound states were not local minima on the 
manifold of constant charge.  This result is quite different from what 
happens in the NLSE where the bound states are local minima on the manifold of 
constant mass. So one cannot assume that the sign
found in Eq. \eqref{second} yields information about the stability of the 
solution.  On the other hand we can assume by continuity that there is a 
region where the analysis of stability in the mNLSE will give us information 
about stability at least in the regime where the expansion parameter
$\epsilon/2m$ is small. For the mNLSE we can use the scaling argument or the 
auxiliary variational approach to discuss stability. It is interesting that 
Derrick \cite{ref:derrick} in his seminal paper was unable to find a suitable 
method for discussing stability for self-interacting spinor theories. 

For the mNLSE the Hamiltonian for the S-S interactions is given by 
\bq
H = \int \frac{dx}{2m} \left[ \psi^\star_x \psi_x\left(1+ \frac{g^2}{2m} (\psi^\star \psi)^k \right) \right] 
-\frac{g^2}{k+1} (\psi^\star \psi)^{k+1}. 
\eq
It is well known that using stability with respect to scale transformation to 
understand domains of stability applies to this type of Hamiltonian. 
This  Hamiltonian is a sum of two positive and one negative term i.e.
\bq 
H= H_1+H_2- H_3 \>.
\eq
For the V-V case, the Hamiltonian is instead
\bq
H=H_1-H_2 -H_3 \>.
\eq
We also know that $H_2$ is of order $g^2/2m$ and is presumed small. 
If we again make a scale transformation on the solution which preserves the 
mass $M = \int  \psi^\star \psi dx$,
\bq
\psi_\beta = \beta^{1/2} \psi(\beta x) \>,
\eq
we obtain
\bq
H=  \beta^2  H_1\pm \beta^{2+k} H_2-  \beta^k H_3 \>.
\eq
Here the upper(lower) sign corresponds to the S-S (V-V) case. 
The first derivative is:
\bq
\frac{\partial H}{\partial \beta} = 2 \beta H_1 \pm (2+k) \beta^{k+1} H_2 - k \beta^{ k-1}  H_3 \>.
\eq
Setting the derivative to zero at $\beta =1 $ gives the equation consistent 
with the equations of motion:
\bq
k H_3 = 2 H_1 \pm (2+k)H_2 \>.
\eq
The second derivative at $\beta =1$ can now be written as
\bq
\frac{\partial^2 H}{ \partial \beta^2} = (4-2k) H_1 \pm  2 (2+k) H_2 \>.
\eq
This will be positive for $k <2$ and the addition of a 
small $H_2$ should extend the stability of the solutions beyond $k=2$  in the 
S-S case. However, in the V-V case there is a somewhat lower region of stability.
At $k=2$, as we shall see below, the
 usual NLSE solitary waves blow up once the mass exceeds a critical value.
For $k=1$, numerical experiments for the time evolution of an initial wave 
of the form  
\bq
\psi(x, t=0) = \sqrt{\beta/2} \sech (\beta y) e^{i ( mvx - \frac{1}{2} mv^2t -\epsilon_0 t ) }
\eq 
at $t=0$  relaxed to an exact solitary wave solution of the mNLSE that  was not very different than the NLSE solution \cite{ref:Toyama}.  This result supports the conclusion that the solitary waves of the mNLSE are stable for k=1.


\section{Self-similar Analysis of  blowup and critical mass for the NLSE and the mNLSE}

To study in a ``mean field" approximation blowup and critical mass, we look for self-similar solutions of the form: 
\bq  
\psi (x,t) = A(t) f (\beta y)   \exp i \left[ m v y + \Lambda(t)  y^2+\omega t \right] \>.
 \eq
Here $\Lambda(t), A(t)$ and~$\beta(t)$ are arbitrary functions of time alone, and  $y=x-vt$.  What we have in mind is to start at $t=0$ with the exact solution of the form $ A \, \sech^{1/k}  (D y)$ and assume that this solution just changes during the time evolution  in amplitude and width conserving mass. With this assumption one can derive the dynamical equations for $A$ and~$D$ from the action principle. 
The action for the NLSE is given by 
\bq  
\Gamma = \int dt L \>,  \label{eq:gamma-definition}
\eq
where L is given by
\bq  
L = {\frac{i}{2}} \int d^d x (\pstar \psi_{t} -\pstar_{t}\psi) - H  \>, \label{eq:action1}
\eq
with
\bq  
H = \int d x \left [ \frac{\pstar_x \psi_x}{2m} -
g^2{\frac{(\pstar\psi)^{k+1}}  {k+1}} \right ] \>.
\eq
The NLSE  follows from the Hamilton's principle of least action:
\bq  {\frac{\delta\Gamma}{\delta\psi}} =
{\frac{\delta\Gamma}{\delta\pstar}} = 0 \>.
\eq

The NLSE has three conservation laws: mass, momentum and energy which can be 
derived from Noether's theorem in the usual fashion. The conservation of mass
\bq  
M = \int \pstar \psi dx = \frac{A^2}{\beta} C_1,~~~ C_1 = \int_{-\infty}^{\infty } f^2(z) d z \>,
\eq
allows one to rewrite $A(t)$ in terms of the conserved mass and the width  
parameter $\beta$ and a constant $C_1$ whose value depends on $f(z)$.  Thus, 
\bq  
A^2 = \frac {M \beta}{C_1} \>.
\eq
For $f(z) = \sech^\gamma(z)$, one obtains 
\bq
C_1= \frac{\sqrt{\pi } \Gamma (\gamma )}{\Gamma \left(\gamma
   +\frac{1}{2}\right)} \>.
   \eq
First consider the kinetic energy (KE) term in the Lagrangian Density
\bq  
 {\frac{i}{2}}  (\pstar \psi_{t} -\psi_{t}^\ast \psi) = f^2 \frac {M \beta}{C_1} \left[m v^2  - {\dot \Lambda} y^2 + 2  v \Lambda y  - \omega  \right] \>.
\eq
Integrating over space and scaling out $\beta$, we obtain
\bq
KE/M = m v^2 -\omega - {\dot \Lambda} G^2 \frac{C_2}{C_1}  \>,
\eq
where  $G= \frac{1}{\beta} $ and 
\begin{align}
C_2  = & \int_{-\infty}^{\infty}  z^2 f ^2 (z) d z 
\\ \nonumber
          = & \frac{2}{\gamma ^3} \, 4^{\gamma -1} \,  _4F_3(\gamma ,\gamma ,\gamma ,2 \gamma ;\gamma +1,\gamma +1,\gamma +1;-1) \>.
\end{align}
Next consider 
\bq
H_0 = \int dx \frac{1}{2m}   \frac{\partial \pstar}{\partial x} \frac{\partial \psi}{\partial x} \>.
\eq
We obtain 
\ba
H_0/M&& = \frac{m v^2}{2 }  +  \frac{C_3}{C_1}   \frac{1}{2m G^2 }  +  4  \Lambda^2  \frac{C_2}{C_1}  \frac{G^2}{2m}  \>,
\ea
where
\bq
C_3  = \int_{-\infty}^{\infty }( f^\prime)^2 (z) d z =\frac{\sqrt{\pi } \gamma  \Gamma (\gamma +1)}{2 \Gamma \left(\gamma
   +\frac{3}{2}\right)} \>. \eq
Finally for the interaction term:
\bq
H_I = - \frac{g^2}{k+1} \int dx( \psi^\ast \psi)^{k+1} \>,
\eq
we obtain
\bq
H_I/M =    - \frac{g^2}{(k+1)}\frac{C_4}{C_1}\left( \frac{M }{C_1 G}\right)^k \>,
\eq
where
\bq
C_4=   \int_{-\infty}^{\infty }f^{(2k+2)} (z) d z   =\frac{\sqrt{\pi } \Gamma [(k+1) \gamma ]}{\Gamma [(k+1) \gamma
   +\frac{1}{2}]} \>.
\eq
Putting this together we get the following ``effective Lagrangian" for the 
time dependent functions $G, \Lambda$:
\begin{align}
L = & \frac{mv^2}{2} - \omega - {\dot \Lambda} G^2 \frac{C_2}{C_1}  - \frac{C_3}{C_1}   \frac{1}{2m G^2 } 
\\ \notag  
    = & - 4  \Lambda^2  \frac{C_2}{C_1}  \frac{G^2}{2m}-  \frac{g^2}{(k+1)}\frac{C_4}{C_1} \Bigl ( \frac{M }{C_1 G} \Bigr )^k \>.
\end{align}
Lagrange's equation for $\Lambda$ yields
\bq
 \Lambda = \frac { 2 m \dot G } {4 G}  \label{lambda} \>.
\eq
The first integral of the second order differential  equation resulting 
from the Lagrange's equation for $G$ can be obtained by setting the 
conserved Hamiltonian to a constant $E$.
One then has
\bq
E =  \frac{C_3}{C_1}   \frac{1}{2m G^2 }  + 4  \Lambda^2  \frac{C_2}{C_1}  \frac{G^2}{2m} - \frac{g^2}{(k+1)}\frac{C_4}{C_1} \Bigl ( \frac{M }{C_1 G} \Bigr )^k \>.
\eq
Using  Eq. \eqref{lambda} we obtain the first order differential equation 
for $G$:
\bq
E =    \frac{C_2}{C_1}  \frac{ 2m \dot G ^2}{4}+\frac{C_3}{C_1}   \frac{1}{2m G^2 }   - \frac{g^2}{(k+1)}\frac{C_4}{C_1}\Bigl ( \frac{M }{C_1 G} \Bigr )^k \>.
\eq 
We notice that at the critical value of  $k =2$, that the last two terms both 
go like $1/G^2$. Self-focusing occurs when the width
can go to zero.  Since $\dot G^2 $ needs to be positive, this means that at 
$k=2$, the mass has to be greater than or equal to  $M^\ast$
for $G$ to be able to go to zero. Here
\bq
\frac{g^2}{3}\Bigl ( \frac{M^\star }{C_1} \Bigr )^2 = \frac{C_3}{C_4}   \frac{1}{2m } \>,
\eq
or 
\bq
  \sqrt{2m} g M^\star = \sqrt{ \frac{3 C_1^2 C_3}{C_4} }=  \frac{\pi}{3}  \sqrt{2} =  2.7207 \ldots \>,
 \eq
provided we use  the exact solution for $k=2$, namely  $f=\sech^{1/2}(z)$ 
(which is a zero-energy solution). This  agrees well with numerical 
estimates of the critical mass  \cite{numerical} and is slightly lower than 
the variational estimate obtained earlier by 
Cooper \emph{et al.} \cite{variational} using post-Gaussian trial wave 
functions. In the supercritical case we have that
\bq
 \frac{C_2}{C_1}  \frac{ 2m \dot G ^2}{4}= \frac{g^2}{(k+1)}\frac{C_4}{C_1}\Bigl ( \frac{M }{C_1 G} \Bigr )^k \>.
\eq 
Thus $G$ approaches zero in a finite time in this self-similar 
approximation with critical index: 
\bq
G  \approx (t-t_c)^{2/(k+2)} \>.
\eq
This ``mean-field" result was obtained earlier in \cite{cooper3,variational}. 

Now we would like to see how this argument is modified when we add the 
$\frac{1}{2m}$ corrections coming from the non-relativistic reduction of the 
NLDE. We now have:
\bq  
L = {\frac{i}{2}} \int d x (\pstar \psi_{t} -\pstar_{t}\psi) - H  \>, \label{eq:action2}
\eq
where for the mNLSE, the Hamiltonian is given by 
\begin{align}
H = & \int dx  \frac{1}{2m} \left[ \psi^\star_x \psi_x\left(1\pm \frac{g^2}{2m} (\psi^\star \psi)^k \right) \right] 
\\ \nonumber
        & -\frac{g^2}{k+1} (\psi^\star \psi)^{k+1} \>. 
\end{align}
Here upper (lower) sign corresponds to the S-S (V-V) case.
Now we get one more term in the energy conservation equation. Also 
Lagrange's equation for $\Lambda$ gets modified.
The new term is 
\begin{align}
\delta H/M = & \pm \frac{1}{M} \frac{g^2}{4m^2} \int dx   \psi^\star_x \psi_x\ (\psi^\star \psi)^k  
\\ \nonumber 
= & \pm \frac{g^2}{4m^2} \left( \frac{M}{C_1}\right)^k 
\\ \nonumber 
& \times \left [  \frac{E_1}{C_1} G^{-(k+2)} + \frac{C_2}{C_1} m^2 v^2 G^{-k} + \frac{E_2}{C_1} 4 \Lambda^2 G^{2-k} \right ] \>, 
\end{align}
where
\begin{align}
E_1 = & \int_{-\infty}^{\infty}( f^\prime)^2 f^{(2k+2)} (z) d z  = \frac{\sqrt{\pi } \gamma ^2 \Gamma [(k+2) \gamma ]}{2 \Gamma [(k+2)
   \gamma +\frac{3}{2}]}\nonumber \\
\end{align}
and
\begin{widetext}
\begin{align}
E_2 =  \int_{-\infty}^{\infty} z^2 f^{(2k+2)} (z) d z  
=  \frac{2^{2 (k+1) \gamma -1}}{(k+1)^3 \gamma ^3}
\, _4F_3(k \gamma +\gamma ,k \gamma +\gamma
   ,k \gamma +\gamma ,2 k \gamma +2 \gamma ;k \gamma +\gamma +1,k \gamma
   +\gamma +1,k \gamma +\gamma +1;-1)
   \>.
\end{align}
\end{widetext}
Lagrangian's equation for $\Lambda$ now yields
\bq
\Lambda =  2m \frac{\dot G}{ 4G} \left[ 1\pm \frac {g^2}{2m} \left( \frac{M}{C_1}\right)^k  \frac{E_2}{C_2} G^{-k} \right]^{-1} \>.
\eq
Conservation of energy in the comoving frame ($v=0$) now leads to 
\ba
&&E =  \frac{C_3}{C_1}   \frac{1}{2m G^2 }  + 4  \Lambda^2  \frac{C_2}{C_1}  \frac{G^2}{2m} - \frac{g^2}{(k+1)}\frac{C_4}{C_1}\left( \frac{M }{C_1 G}\right)^k  \nonumber \\
&&  \pm \frac{g^2}{4m^2} \left( \frac{M}{C_1}\right)^k \left[  \frac{E_1}{C_1} G^{-(k+2)} 
 + \frac{E_2}{C_1} 4 \Lambda^2 G^{2-k} \right] \>,
\ea
or
\begin{align}
E =  & \frac{C_3}{C_1}   \frac{1}{2m G^2 } \left[1\pm \frac{g^2}{2m} \frac{E_1}{C_3} \left( \frac{M}{C_1G }\right)^k  \right] 
\nonumber \\
& +\frac{2m}{4} \dot G^2 \frac{C_2}{C_1} \left[1\pm \frac{g^2}{2m} \frac{E_2}{C_2}  \left( \frac{M}{C_1G }\right)^k \right]^{-1} 
\nonumber \\ 
& - \frac{g^2}{(k+1)}\frac{C_4}{C_1} \left( \frac{M }{C_1 G}\right)^k  \nonumber \>.
\end{align}
From this expression we again see that $k=2$ is the critical value. If the 
initial value of $G$ is large enough so we can ignore the
$g^2/2m$ corrections then in order for $\dot G^2 >0$, so that the width can 
decrease, one needs that 
\bq
  \sqrt{2m} g M^\star \geq  \sqrt{ \frac{3 C_1^2 C_3}{C_4} } \>.
\eq
When $G$ gets very small then the $g^2/2m$ corrections get large and our 
expansion breaks down.  Blowup then needs to be studied
using the full NLDE.  We intend to do numerical studies of blowup in the 
NLDE in the near future.

 \section{Conclusions}
In this paper we have found new solutions to the NLDE with arbitrary 
nonlinearity parameter $k$ in the case of both the S-S and V-V interactions. 
The solutions for the S-S interactions have the property that 
for $\omega  > \omega_c(k)$ the shape of the solitary wave is similar to 
a $\sech^\gamma(x)$ profile, whereas for $\omega \le \omega_c(k)$, the shape 
is double humped.  In the V-V case, the shape of the profile is always of 
the form $\sech^\gamma(x)$. We discussed the nonrelativistic reduction of the 
NLDE and obtained a modified NLSE (mNLSE)  whose stability properties could 
be studied in a variety of ways.  By continuity we expect that at least in 
the regime where the solutions of the NLDE are small perturbations of those 
of the NLSE, the solutions we have found will be stable for $k<2$.  
We discussed the case $k=2$ for the mNLSE approximation in detail as well 
as blowup for $k >2$ using a self-similar ansatz.  

Before ending we point out some of the possible open questions.

\begin{enumerate}

\item Is there a connection between instability and the double hump behavior?

\item  In the V-V case we notice from Fig. 4 that while for $k<2$,
$\omega(g_{min})>\omega(g_{max})$, for $k>2$, the opposite is true. Is
this somehow related to the fact that the NLDE V-V bound states are stable
(unstable) for $k < (>) 2$? Further, the dip in the value of $\omega(g_{min})$
precisely occurs around $k=2$ in both the S-S and the V-V cases. Is that just
a coincidence or is it related to the instability for $k >2$?

\item For $k=1$, it is known that the bound states of N localized fermions
are stable in both the S-S and V-V cases. It would be interesting to
examine if this continues to be true for arbitrary positive $k$. 

\end{enumerate}

We hope to address some of these questions in the near future.
Also we intend to do numerical simulations of collisions to see how energy 
and charge are exchanged, and also study blowup to understand whether there 
is much difference between self-focusing in the NLDE and the NLSE.  

\begin{acknowledgments}
This work was performed in part under the auspices of the U.S.  
Department of Energy.  F.C. and B.M. would like to thank the Santa Fe Institute 
for its hospitality during the completion of this work. A.K. would like to
thank Center for Nonlinear Studies, Los Alamos National Laboratory, for warm 
hospitality during his stay. F.C. would like to thank T.~Goldman for useful 
conversations about the nonrelativistic reduction of the Dirac equation. 
\end{acknowledgments}

\end{document}